\documentclass[aps,prl,twocolumn,nopacs,superscriptaddress,groupedaddress]{revtex4}  
\usepackage{graphicx,amssymb,epsfig,amsmath}
\usepackage{epstopdf}
\usepackage{dcolumn}   
\usepackage{bm}        
\usepackage[loose]{units}
\usepackage{color}
\usepackage{ulem}
\begin{document}

\title{Time-Resolved Magnetic Relaxation of a Nanomagnet on Subnanosecond Time Scales}

\author{H. Liu}
\author{D. Bedau}
\email{db137@nyu.edu}
\affiliation{Department of Physics, New York University, New
             York, NY 10003, USA}
\author{J. Z. Sun}
\affiliation{IBM T. J. Watson Research Center, P.O. Box 218, Yorktown Heights, New York, 10598 USA}
\author{S. Mangin}
\affiliation{Institute Jean Lamour, UMR CNRS 7198, Nancy Universit\'{e},
             Vandoeuvre, France}
\author{E.~E. Fullerton}
\affiliation{CMRR, University of California at San Diego,
             La Jolla, CA 92093, USA}
\author{J.~A. Katine}
\affiliation{San Jose Research Center, Hitachi-GST,
             San Jose, CA 95135, USA}
\author{A.~D. Kent}
\affiliation{Department of Physics, New York University, New
             York, NY 10003, USA}

\date{April 8, 2012}

\begin{abstract} 
We present a two-current-pulse temporal correlation experiment to study the
intrinsic subnanosecond nonequilibrium magnetic dynamics of a nanomagnet during
and following a pulse excitation. This method is applied to a model spin-transfer
system, a spin valve nanopillar with perpendicular magnetic anisotropy.
Two-pulses separated by a short delay ($<500$ ps) are shown to lead to the same
switching probability as a single pulse with a duration that depends on the
delay. This demonstrates a remarkable symmetry between magnetic excitation and
relaxation and provides a direct measurement of the magnetic relaxation time. The
results are consistent with a simple finite temperature Fokker-Planck macrospin
model of the dynamics, suggesting more coherent magnetization dynamics in this
short time nonequilibrium limit than near equilibrium.
\end{abstract}

\maketitle 

The control of magnetization on short timescales has become an area of intense
research and many methods are used to excite a magnetic system, including
spin-currents~\cite{Koch2004,Garzon2008}, magnetic fields~\cite{Schumacher2003}
and optical pulses~\cite{Vaterlaus1991,Beaurepaire1996}. It also has recently
become possible to drive individual nanometer scale magnetic elements far from
equilibrium using spin-currents and probe their dynamical response. However, most
electronic transport studies focus on the excitation during the drive pulses
\cite{Devolder_single-shot_2008,cui:single-shot:2010}. Less is known about the
relaxation processes following a pulse excitation. Yet, this is very important
for fundamental and practical reasons. First, the time scales of the relaxation
far from equilibrium are not known and may differ from those determined in
experiments that probe low amplitude magnetic excitations, such as ferromagnetic
resonance. Second, these time scales determine the speed at which nanomagnets can
be written and read in magnetic memories, including in spin-transfer torque
magnetic random access memories (STT-MRAM).

It is now well known that spin-polarized currents and spin-transfer torques
(STTs) can reverse the direction of the magnetization on subnanosecond timescales
\cite{Koch2004,bedau:ultrafast:2010}. In fact, many experimental methods have
been developed to study spin-transfer switching in both spin valves and magnetic
tunnel junctions (MTJs). While MTJs are promising for applications because of
their large magnetoresistive (MR) readout signals ($>100 $\%), all-metallic spin
valves permit one to apply much larger currents and thus can be driven farther
away from equilibrium. Direct single-shot time-resolved electrical measurements
have been carried out in MTJs~\cite{Devolder_single-shot_2008,
cui:single-shot:2010}, but have not been reported in spin valve nanopillars,
because of their low impedance and small magnetoresistance (MR) ($\lesssim 5$\%),
which results in their switching signals being to small for nanosecond
measurements. Further, time-resolved STT measurements are insensitive to the
magnetization dynamics after the excitation, as the excitation current is the
source of the electrical readout signal. Therefore, to time-resolve the
relaxation dynamics, a different experimental method is needed.

In this paper we present an all-electrical two-pulse correlation method that
yields quantitative information on the form and timescales of magnetic excitation
and relaxation, with \unit[50]{ps} time resolution. This method is applied to a
model spin-transfer system, a spin valve nanopillar with perpendicular magnetic
anisotropy. Two-pulses separated by a short delay ($<500$ ps) are shown to lead
to the same switching probability as a single pulse with a duration that depends
on the delay--demonstrating a remarkable symmetry between magnetic excitation and
relaxation--and providing a direct measurement of the magnetic relaxation time.
The observed symmetry is shown to be consistent with a finite temperature
macrospin model.

\begin{figure*}[t] \centering \includegraphics[width=0.8\textwidth]{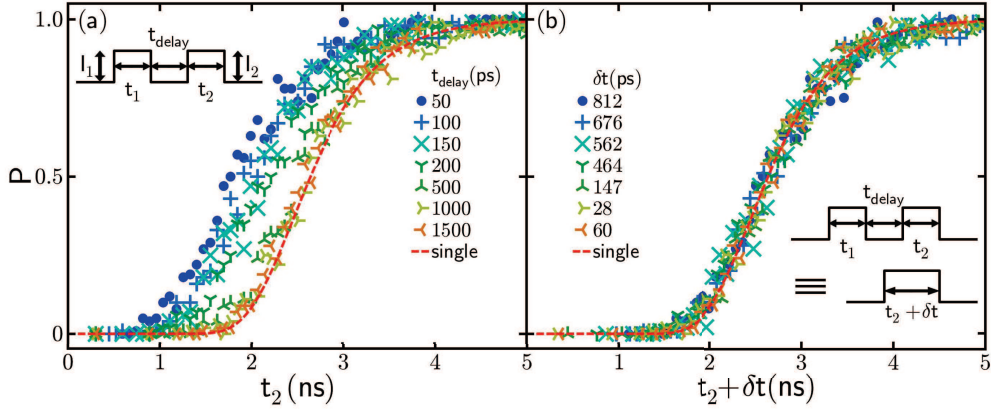}
\caption{(a) The total switching probability, $P_{AP \to P}^\mathrm{double}$, as
a function of the second pulse duration $t_2$, for a current $I_1$ of
\unit[8.5]{mA} and different delays. The dashed line is the switching probability
$P_{AP \to P}^{(2)}$ for a single \unit[8.5]{mA} pulse of duration $t_2$. (b) The
same data is plotted with shifted time axis, with the shift dependent on the
delay $\delta t(t_{\mathrm{delay}}$). The data closely overlaps showing that the
switching probability of a double pulse ($t_1 + t_{\mathrm{delay}} + t_2$) is the
same as that of a single pulse with a longer duration, i.e. $t_2 + \delta t
(t_{\mathrm{delay}})$, as illustrated schematically by the pulses drawn within
the figure.}
\label{fig:1}
\end{figure*}

The system we studied has two stable magnetic states $A$ and $B$, which are
nearly degenerate in equilibrium. We are interested in how the system relaxes to
these states after being excited by a spin-polarized current. The basic idea of
our method is to compare the switching probability of two pulses separated by a
delay much longer than the magnetic relaxation time to that of a composite pulse,
i.e. two pulses separated by a short delay. With a composite pulse, the first
pulse alters the magnetic state on a time scale that changes the dynamics excited
by the second pulse. By choosing the amplitudes and polarities of the first and
second pulse we can study different aspects of the relaxation processes. For
example, to study the relaxation back to the initial state (i.e. state $A$) after
a pulse, we can apply two pulses with identical polarities that both would be of
the appropriate polarity to switch the sample from state $A$ to $B$. If the delay
is much longer than the relaxation time $t_{\mathrm{relax}}$, the two pulses can
be considered to be independent events and the total probability of switching
from state $A$ to state $B$ is:
\begin{eqnarray}
P_{A \to B}^\mathrm{double}\left(t_{\mathrm{delay}} \gg t_{\mathrm{relax}}
\right) & = &P_{A \to B}^{(1)} \\ & + & \left(1 - P_{A \to B}^{(1)} \right)
\times P_{A \to B}^{(2)} \nonumber
\label{eq:chainrule}
\end{eqnarray}
Here $P_{A \to B}^{(1)}$, $P_{A \to B}^{(2)}$ and $P_{A \to B}^\mathrm{double}$
are the switching probabilities from state $A$ to $B$ for the first, the second
and for both pulses combined and we have assumed no reverse switching during the
measurement, i.e. $P_{B \to A}=0$. If $t_{\mathrm{delay}}$ is within the
relaxation time of the system, the total switching probability will increase
compared to Eq.~\eqref{eq:chainrule}:
\begin{equation}
P_{A \to B}^\mathrm{double}\left(t_{\mathrm{delay}} < t_{\mathrm{relax}} \right)
 > P_{A \to B}^\mathrm{double}\left(t_{\mathrm{delay}} \gg t_{\mathrm{relax}}
\right)
\label{eq:short_delay}
\end{equation}
Using these equations we can resolve the relaxation dynamics of the sample by
measuring and comparing the switching probabilities as a function of
$t_{\mathrm{delay}}$ \cite{Eq2}.

Our sample is a spin-valve nanopillar consisting of a Ni\textbar Co free layer
and exchange coupled Ni\textbar Co and Co\textbar Pt multilayers as the reference
layer, both of which have perpendicular magnetization anisotropy (PMA). (For the
full layer stack see \cite{Bedau:Spin:2010, bedau:ultrafast:2010}.) Such PMA
samples have a simple uniaxial anisotropy energy landscape that permits a direct
comparison to theoretical models. The sample presented in this work is
$\unit[100]{nm}\times\unit[100]{nm}$ in size, fabricated by e-beam and optical
lithography \cite{Mangin2006,Mangin2009}. We have studied more that $20$ samples
of different shapes and sizes and found similar results. The equilibrium states
are antiparallel ($A \equiv$ AP) and parallel ($B \equiv$ P) magnetization
alignment of the free and reference layers and there is $0.3\%$ MR, well within
the accuracy of our method. The room temperature coercive field is
$\unit[100]{mT}$, large enough so that no reverse switching occurs
\cite{min:back:hopping:2009}. The zero temperature switching current for
antiparallel (AP) to parallel (P) switching is $I_{c0} = \unit[6.5]{mA}$.

We use an arbitrary waveform generator to apply two pulses separated by a time
delay $t_{\mathrm{delay}}$, as illustrated by the inset of Fig.~\ref{fig:1}(a).
The state of the spin valve is determined by a resistance measurement using a
lock-in amplifier both before and after the injection of the double-current
pulse. We note that the lock-in current is small enough (\unit[300]{$\mu$A}) not
to affect the spin valves and not to induce reverse switching
\cite{Bedau:Spin:2010}. All experiments described here were conducted at room
temperature.

Initially the sample was brought into the AP state by the application of an
easy-axis magnetic field of appropriate direction and magnitude. A short current
pulse of positive polarity $I_1 = \unit[8.5]{mA}$ and $t_1 = \unit[0.9]{ns}$ was
then applied. A longer pulse of positive polarity would eventually switch the
sample into the P state (as $I_1>I_{c0}$). However, the duration has been chosen
such that the switching probability is nearly zero ($\lesssim 1/100$), as shown
by the dashed curve (i.e. the single pulse curve) in Fig. \ref{fig:1}(a) at
\unit[0.9]{ns}. After a variable delay from \unit[50]{ps} to \unit[1.5]{ns}, a
second pulse with the same amplitude ($I_2 = I_1$) and a variable duration $t_2$,
from \unit[50]{ps} to \unit[5]{ns}, was applied. The total switching probability
distribution $P_{AP \to P}^\mathrm{double}$ was then measured by repeating the
process $100$ to $10\,000$ times for each pulse set.

As seen in Fig.~\ref{fig:1}(a), for delays longer than \unit[500]{ps} the
switching probability distribution is the same as that of the second pulse alone
$P_{A \to B}^\mathrm{double} = P_{AP \to P}^{(2)}$, which is consistent with
Eq.~\eqref{eq:chainrule} for $P_{A \to B}^{(1)} = 0$. This indicates that these
delays are longer than the magnetization relaxation time. The magnetization state
excited by a \unit[0.9]{ns} current pulse decays to equilibrium in under
\unit[500]{ps}, within our experimental accuracy. However, for delays less than
\unit[500]{ps}, the switching probability $P_{AP \to P}^\mathrm{double}$ is
larger than that of just a single pulse.

Remarkably, even with a composite pulse, the switching probability versus the
second pulse duration is seen to follow the same functional form as that for a
single pulse. This is shown in Fig.~\ref{fig:1}(b), where the probability data,
shifted by an amount $\delta t(t_{\mathrm{delay}})$ -- a quantity that depends on
the delay between the pulses -- is seen to closely overlap. This clearly
demonstrates that the switching probability of a composite pulse ($t_1 +
t_{\mathrm{delay}} + t_2$) is equal to that of a single pulse with an increased
duration ($t_2 + \delta t$). Since this is satisfied for all $t_2$, a natural
conclusion is that the ensemble averaged state of the sample after the first
pulse and the delay ($t_1 + t_{\mathrm{delay}}$) is the same as that after a
single pulse with a duration of $\delta t$, as illustrated schematically by the
pulse shapes in the inset of Fig.~\ref{fig:1}(b).

\begin{figure}[h] \centering \includegraphics[width=0.8\linewidth]{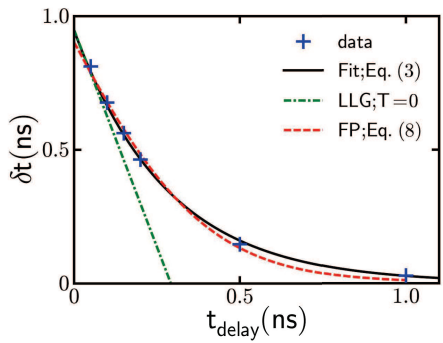}
\caption{$\delta t$ versus delay $t_{\mathrm{delay}}$. The blue crosses are
determined from the experimental data in Fig.~\ref{fig:1}. The black solid curve
is an exponential fit to Eq.~\eqref{eq:fitting}. The green dash-dotted line is a
calculation from the LLG equation for $T = 0$. The red dashed curve is a
calculation based on a Fokker-Planck Analysis (Eq.~\eqref{eq:FP}).}
\label{fig:2}
\end{figure}

From the data shown in Fig.~\ref{fig:1} we determine $\delta t$ as a function of
the delay $t_{\mathrm{delay}}$. This is shown as crosses in Fig.~\ref{fig:2}. The
black solid curve is an exponential fit to the data where $\tau_L$ is the
lifetime of the excitation:
\begin{equation}
  \delta t = t_0 \exp \left(-\frac{t_{\mathrm{delay}}}{\tau_L}\right)
\label{eq:fitting}
\end{equation}
with the parameters $t_0= \unit[0.95]{ns}$ and $\tau_L =\unit[0.28]{ns}$.
For no delay, $t_{\mathrm delay}=0$, clearly $\delta t=t_1$ and thus $t_0
=t_1$, which is indeed fulfilled within the time resolution of the
experiment ($\simeq 50$ ps). Eq.~\eqref{eq:fitting} links the duration of the
excitation, the delay and the corresponding equivalent pulse duration, which is
related to the magnetization relaxation rate.

The data in Fig.~\ref{fig:2} is unexpected at first, as a zero temperature
macrospin Landau-Lifshitz-Gilbert (LLG) model predicts a linear relationship
between $\delta t$ and $t_{\mathrm delay}$, as shown by the green dash-dotted
line in Fig.~\ref{fig:2}. This linear relationship follows from the LLG equation.
For a uniaxial nanomagnet in zero applied magnetic field, the polar angle of the
magnetization $\theta$, i.e. the angle between the magnetization and the easy
axis, satisfies \cite{Sun2000}:
\begin{equation}
\frac{\displaystyle d \theta}{\displaystyle d t} = \frac{\displaystyle
1}{\displaystyle \tau_D} \left(\frac{\displaystyle I}{\displaystyle
I_{c0}} - \cos \theta \right)\sin \theta
\label{eq:llg_theta}
\end{equation}
with:
\begin{equation}
\tau_D = \left(\frac{\displaystyle 1 + \alpha ^ 2}{\displaystyle \alpha
\gamma \mu_0 H_k} \right)
\label{eq:tau_D}
\end{equation}
\begin{equation}
I_{c0} = \frac{\displaystyle 2 e M_s V \alpha}{\hbar \eta} \mu_0 H_k
\label{eq:Ic0}
\end{equation}
$\tau_D$ is the relaxation time due to damping and depends on the material
parameters, the anisotropy field $H_k$, the Gilbert damping parameter $\alpha$
and the gyromagnetic ratio $\gamma$. $\tau_D$ is typically hundreds of
picoseconds. $I_{c0}$ is the zero temperature critical current as we mentioned
earlier, representing the threshold current for switching. It depends on the
magnetization density $M_s$, the nanomagnet's volume $V$ and the spin
polarization of the current $\eta$. From Eq.~\eqref{eq:llg_theta} both excitation
($I > I_{c0}$) and relaxation ($I = 0$) follow the same functional form for small
angles $\theta$ and are only different in their respective time constants. It is
straightforward to show that the duration of a single pulse ($\delta t$), which
brings the sample to the same state as the first pulse ($t_0$) followed by the
delay ($t_{\mathrm{delay}}$) satisfies:
\begin{equation}\label{eq:msdeltat}
  \delta t = t_0 - \left( \frac{I}{I_{c0}} -1 \right)^{-1} t_{\mathrm{delay}}
\end{equation}
Hence the initial slope of $\delta t$ versus $t_{\mathrm{delay}}$, shown as the
green dashed dotted curve in Fig.~\ref{fig:2}, depends on the current
overdrive, $I/I_{c0} -1$, and provides an independent method to determine the
zero temperature critical current. However, for long delays this equation is
unphysical, as it predicts a negative $\delta t$.

A physical picture is that during the first pulse the magnetization evolves from
a thermal distribution of initial angles near the north pole ($\langle \theta_0
\rangle \sim \sqrt{\pi k_B T/ 4U} \approx 6^\circ$, where $U$ is the energy
barrier and $k_B T$ is the thermal energy) to a larger angle $\theta_1$. During
the delay, and provided  $\theta_1<90^\circ$, the polar angle decays back toward
the north pole. However, in this zero temperature model the magnetization decays
back to $\theta_2=0$, i.e. to an angle that can be less than the initial angle,
$\theta_0$.

\begin{figure*}[t] \centering \includegraphics[width=0.9\linewidth]{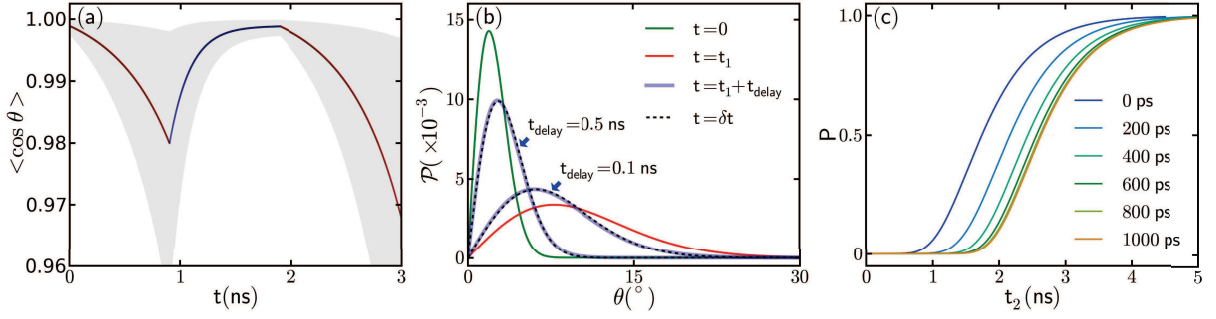}
\caption{Results of a Fokker-Planck calculation. (a) Ensemble averaged
magnetization projection versus time for a \unit[1]{ns} delay. During the first
pulse the average magnetization projection increases with time as shown by the
first red curve, and it decreases with time during the delay as shown by the blue
curve. The shaded region depicts the angular range containing $10\%$ to $90\%$ of
the states. (b) Occupation probability plotted as a function of the polar angle.
The green curve is the initial thermal equilibrium Boltzmann distributed polar
angle. The red solid curve is the probability distribution at the end of the
first pulse. The light blue solid curve is the probability distribution after the
first pulse and delay, $t_{\mathrm{delay}} = 0.1$ and $0.5$ ns. The black dashed
curve is the probability distribution corresponding to a pulse of duration
$\delta t$. The corresponding solid light blue and black dashed curves match well
showing that these distributions are nearly same -- that the probability
distribution after a pulse and delay is nearly the same as that after a single
shorter pulse $\delta t$ -- as inferred from the scaling of the experimental data
presented in Fig.~1(b). (c) Switching probability distributions plotted as a
function of the second pulse duration for different delay times. As in the
experiments, the probability distributions have the same form but are shifted
horizontally, along the time axis.}
\label{fig:FP}
\end{figure*}

The difference between the LLG model prediction and the experimental results can
be explained by thermal fluctuations both during the excitation and relaxation
processes. As is well known, thermal fluctuations lead to a spread of the initial
angles of magnetization about the north pole. A spin-current pulse works to
rotate the magnetization away from the easy axis direction and, in essence,
amplifies these magnetization fluctuations. Once the magnetization deviates from
the easy axis direction, the magnetization polar angle grows exponentially.
However, thermal noise also slows down the relaxation from a simple exponential
decay process predicted by the LLG equations and leads to relaxation to a thermal
distribution. Therefore, thermal fluctuations not only change the timescales of
the excitation and relaxation processes, they also alter their functional forms.

To model the influence of thermal fluctuations, we simulated the double pulse
measurement using the Fokker-Planck Equation \cite{Brown1963}, which has the
following form for a uniaxial macrospin system:
\begin{eqnarray}
\frac{\displaystyle \partial \mathcal{P}}{\displaystyle \partial t} & = & -
\frac{1}{\tau_D} \frac{\partial }{\partial \theta} \left[\left( \frac{I
}{I_{c0}} \sin \theta - \cos \theta\sin \theta +  \frac{ \cot \theta }{2\xi}
\right)\mathcal{P} \right] \nonumber \\ & + & \frac{1}{2\xi\tau_D}\frac{\partial
^ 2 \mathcal{P}}{\partial \theta ^ 2}
\label{eq:FP}
\end{eqnarray}
where $\mathcal{P} \equiv \mathcal{P}\left(\theta, t\right)$ is the probability
density of the magnetization at the angle $\theta$ and the time $t$.  $\xi$ is
the dimensionless energy barrier, $\xi \equiv U/(k_BT)$. For a uniaxial
nanomagnet $U=1/2M_sH_kcV$.

We start with a Boltzmann distribution centered around the easy axis in the AP
state: \[ \mathcal{P}\left(\theta, t = 0\right) = \left\{
  \begin{array}{l l}
    P_0 \exp\left(\xi \cos^2 \theta\right)\sin\theta & \quad 0^\circ \leq \theta
    \leq  90^\circ \\ 0 & \quad  90^\circ < \theta \leq  180^\circ \\ \end{array}
    \right.
\]
Where $P_0$ is a normalization constant so that ${\displaystyle \int^{\pi}_{0}
\mathcal{P}\left(\theta, t = 0\right)\mathrm{d}\theta = 1}$. For the simulation
we use a fixed duration for the first pulse of \unit[0.9]{ns}, the same as in our
experiments and vary the duration of the second pulse for different delays.
By fitting the measured single pulse switching probability distribution
(i.e. the red dashed curve of Fig.~\ref{fig:1}(a)) to Eq.~\eqref{eq:FP}.
we obtain $\xi = 450$ and $\tau_D = \unit[354]{ps}$.

We then can track the ensemble averaged magnetization projection, $\langle \cos
\theta \rangle =\displaystyle \int_{0}^{\pi} \cos \theta \mathcal{P}\left(\theta,
t\right) \mathrm{d} \theta$ as a function of time $t$, as shown in
Fig.~\ref{fig:FP}(a). The shaded region depicts the angular range containing
$10\%$ to $90\%$ of the states. During the first current pulse ($\unit[0]{ns}
\leq t \leq \unit[0.9]{ns}$), the switching process begins. The averaged
magnetization projection decreases while the occupation probability distribution
spreads out, where the majority of the distribution is still centered around
$\theta = 0^\circ$. During the delay ($\unit[0.9]{ns} \leq t \leq
\unit[1.9]{ns}$), the averaged magnetization projection increases while the
occupation probability distribution narrows toward the initial state. Then during
the second pulse ($t \geq \unit[1.9]{ns}$), the averaged magnetization projection
decreases again as the total distribution expands during the beginning part of
the switching process.

We can further check whether a pulse and a delay ($t_1 + t_{\mathrm{delay}}$)
brings the system into the same state as a single $\delta t$ duration pulse.
Fig.~\ref{fig:FP}(b) shows the occupation probability as a function of the polar
angle for different delay times. The distributions for a pulse and delay  $t_1 +
t_{\mathrm{delay}}$ (the solid light blue curves in Fig.~\ref{fig:FP}(b)) overlap
with those of a single $\delta t$ pulse (dashed curves in
Fig.~\ref{fig:FP}(b))--which is consistent with scaling of the experimental data
shown in Fig.~\ref{fig:1}(b).

The Fokker-Planck simulation results also agree quantitatively with the
experimental data. As shown in Fig.~\ref{fig:FP}(c) the calculated switching
probability distributions have the same form independent of $t_{\mathrm{delay}}$
and can be shifted in time to coincide. The results shifts, $\delta t$s, are in
good agreement with the experimental results, as indicated by the red dashed
curve in Fig.~\ref{fig:2}. The difference between the LLG and the Fokker-Planck
analysis clearly shows the influence of the thermal fluctuations, even though the
total process (excitation plus relaxation) lasts less than \unit[2]{ns}.

In summary, we have presented a method ideally suited to study excitation and
relaxation in a magnetic nanostructure. This method only requires the generation
of pulses with short durations and delays and measurement of the switching
probability. It does not require resolving small signal changes at high bandwidth
and it can also be used to study dynamics after a pulse, with no current applied.

Our method provides a direct (model independent) measurement of the relaxation
timescale by considering when two events are correlated. We have used this method
to study excitation and relaxation in a model system--a thin film nanomagnet with
uniaxial anisotropy excited by spin-current pulses. We have shown that a single
pulse followed by a delay leaves a nanomagnet in a state with the same ensemble
average as that of a single pulse with a shorter duration $\delta t$. This
finding is also model-independent. The $\delta t$s are seen to decrease
exponentially, following the form of Eq.~\ref{eq:fitting}, and giving a
relaxation time of $\unit[280]{ps}$. We note that the relaxation time is more
typically obtained experimentally by fitting to an Arrhenius-N\'{e}el's model
with a variable prefactor (see, for example, \cite{Loth2012}). In general, the
decay of the magnetization might be expected to be more complex than a simple
exponential (Eq.~\ref{eq:fitting}), including multiple relaxation pathways with
different timescales.  However, even in such a case the directly measured
relationship between $\delta t$ and $t_{\mathrm{delay}}$ would provide
information that is important for understanding such relaxation processes. Using
Eq.~\eqref{eq:tau_D} with an anisotropy field of $H_k=0.25$ T (see
Ref.~\cite{bedau:ultrafast:2010}) and a gyromagnetic ratio of
$\gamma=1.76\times10^{11}/$(Ts) (i.e. a Land\'e g-facto of $2$), we find a
Gilbert damping of $\alpha = 0.09$, a value that is larger than that determined
from thin film ferromagnetic resonance measurements of the damping ($\alpha =
0.04 $ \cite{Beaujour2007}). The larger damping inferred from the data may be a
consequence of the large-angle excitation of the magnetization, including
non-linear effects \cite{Tiberkevich2007}, or to changes of the magnet's material
properties associated with device nanofabrication \cite{Ozatay2008}.

Interestingly, our results can be understood within a simple macrospin model that
incorporates thermal fluctuations. In fitting our data to this model we obtain a
dimensionless energy barrier of $\xi = 450$, which is within 25\% of that
expected based on the volume, magnetization and anisotropy of the nanomagnet
($\xi=M_sH_kV/(2kT)= 360$). Our previous results, as well as those of many other
groups, have found energy barriers to reversal for long field
~\cite{moritz:_magnetization_2005,Adam2010,Thomson2006,Gopman2012} and long
current pulses  ($>100$ ns) ~\cite{Bedau:Spin:2010,bedau:ultrafast:2010} to be
only a small fraction of that expected for a macrospin, suggesting the reversal
proceeds by subvolume nucleation ~\cite{Sun2011,Bernstein2011}. For similar
samples, we found that the energy barrier to reversal under long current pulses
to be $\simeq \unit[60]{k_BT}$~\cite{bedau:ultrafast:2010}, $1/6$ the macrospin
value. This suggests that the reversal modes in the short-time nonequilibrium
limit are distinct from those in the long-time limit, as the system may not have
time to follow minimum energy paths or be driven from these paths by the spin
transfer torque. The fundamental explanation is an open question. We expect that
the dependency of the effective energy barrier on pulse amplitude can be further
studied by our method with different pulse conditions. Our double pulse method,
which provides access to the intrinsic time scales of the excitation and
relaxation dynamics, can be used in other cases where the readout signal is also
small, such as current induced domain wall motion.

\section*{Acknowledgments} This research was supported at NYU by NSF Grant
DMR-1006575, USARO Grant No. W911NF0710643  as well as the Partner University
Fund (PUF) of the Embassy of France.

\pagebreak

\end{document}